\documentclass[12pt]{article}
\usepackage{epsf}
\title{ {\bf $H^+ \rightarrow W^+ l_i^- l_j^+$  decay
in the two Higgs doublet model}}

\author{\vspace{1cm}\\
        {\bf E. O. Iltan}
        \thanks{E-mail address:
        eiltan@heraklit.physics.metu.edu.tr}\,\,\, and
   {\bf H. Sundu}
        \thanks{E-mail address:
        sundu@.metu.edu.tr}
 \\
        Physics Department, Middle East Technical University \\
        Ankara, Turkey\\}
\date{}
\begin{document}
\setlength{\baselineskip}{24pt}
\maketitle
\setlength{\baselineskip}{7mm}
\begin{abstract}
We study the lepton flavor violating $H^+ \rightarrow W^+ l_i^-
l_j^+$ and the lepton flavor conserving $H^+ \rightarrow W^+ l_i^-
l_i^+$ $(l_i=\tau, l_j=\mu)$ decays in the framework of the general two
Higgs doublet model, the so-called model III. We estimate the decay width 
of LFV (LFC) decay at the order of magnitude of $(10^{-11}-10^{-5})\, GeV$ 
($(10^{-9}-10^{-4}) \,GeV$), for $200 \,GeV\leq m_{H^\pm}\leq 400\, GeV$, 
and the intermediate values of the coupling
$\bar{\xi}^{E}_{N,\tau \mu}\sim \,5 \,GeV$ ($\bar{\xi}^{E}_{N,\tau
\tau}\sim \,30\, GeV$). We observe that the experimental result of
the process under consideration can give comprehensive information
about the physics beyond the standard model and the existing free
parameters.
\end{abstract}
\thispagestyle{empty}
\newpage
\setcounter{page}{1}
%%%
%%%
\section{Introduction}
The charged Higgs boson carries a distinctive signature of the
Higgs sector in the models beyond the standard model (SM), such as
two Higgs doublet model (2HDM), minimal extension of the standard
model (MSSM). Therefore, its discovery will be an evidence of
multi doublet structure of the Higgs sector. In the literature,
the charged Higgs decays have been widely studied.

The charged Higgs production in hadron colliders was studied in
\cite{Diaz} and more systematic calculations of production process
at LHC have been presented in \cite{Bisset}.  At the LHC the
dominant production channel for $H^+$ is $g b \rightarrow H^+ t$.
One expects more than one thousand  events for a Higgs mass
$m_{H^+}=400\, (GeV)$, with an integrated luminosity of $L=100
fb^{-1}$ \cite{Belyaev}. At tevatron, the CDF and D0
collaborations have searched for $H^+$ bosons through the process
$p\bar{p}\rightarrow t \bar{t}$, with at least one of the top
quark decaying via $t\rightarrow H^+ b$. They have excluded the
regions with light $H^+$ \cite{Abe}. At present the model
independent lower limit on the charged Higgs mass is 
$m_{H^+}> 77.4\, (GeV)$ \cite{Andriga}.

The charged Higgs boson decay into tau and neutrino has been
analyzed in \cite{Christove} and in \cite{Barger} and it was shown
that the dominate decay modes of the charged Higgs boson were
$H^+\rightarrow \tau^+ \nu$ and $H^+\rightarrow t \bar{b}$.
However,  the other candidate for the large branching ratio $BR$
is the process $H^+\rightarrow W^+ h^0$, and it has been examined
in \cite{Santos, Moretti}. The analysis in \cite{Santos} was
related to the $H^+\rightarrow W^+ h^0$  decay, in the framework
of the 2HDM, including loop corrections and for some reasonable 
choice of free parameters, those corrections could be as large as 
$\sim 80 \%$ of the tree level result. In \cite{Moretti}, the 
chances of detecting charged Higgs boson of the MSSM at Large 
Hadron Colliders (LHC), in the $W^+ h^0$ mode, have been studied 
and it was concluded that the charged Higgs boson signal overcomes 
the background for optimum $tan\beta$ values, between 2 and 3. 
In \cite{Yang} the above decay has been analyzed  in the MSSM and 
the electroweak (EW) corrections have been obtained. It was observed 
that, for the low $tan\beta$, these corrections caused an enhancement 
at the order of $20\%$.

The work in \cite{Diaz2} is devoted to the decays of the charged Higgs 
boson, including the radiative modes into decays $W^+ \gamma$
and $W^+ Z$, mostly in the framework of the 2HDM and MSSM.
In \cite{Diaz3}, the analysis of $H^+\rightarrow W^+
\gamma$, $H^+\rightarrow W^+ Z$ and $H^+\rightarrow W^+ h^0$
decays has been done in the context of the effective lagrangian extension 
of the 2HDM. In this work the BRs have been obtained at the order 
of magnitude of $10^{-5}, 10^{-1}$ and $O(1)$, respectively.

Lepton flavor violating (LFV) interactions are interesting, since they do 
not exist in the SM and give strong signal about the new physics beyond. 
Such decays have reached great interest at present and the experimental
search has been improved. $H^0\rightarrow \tau\mu$ decay is an
example for LFV decays and it has been studied in \cite{Cotti, Marfatia}. 
In \cite{Cotti} a large $BR$, at the order of magnitude of
$0.1-0.01$, has been estimated in the framework of the 2HDM. In 
\cite{Marfatia} its BR was obtained in the interval $0.001-0.01$ for 
the Higgs mass range $100-160 (GeV)$, for the LFV parameter 
$\kappa_{\mu\tau}=1$.

Our work is devoted to the analysis of the LFV $H^+ \rightarrow W^+
l_i^- l_j^+$ and the lepton flavor conserving (LFC) $H^+
\rightarrow W^+ l_i^- l_i^+$ $(l_i=\tau, l_j=\mu)$ decays in the
framework of the general 2HDM, the so-called model III. The present LFV  
decay exists with the chain processes, $H^+\rightarrow W^+ (h^{0*},
A^{0*})\rightarrow W^+ l_i^- l_j^+$, where $h^0, A^0$ are CP even
neutral Higgs bosons beyond the SM. This decay is rich in the
sense that its decay width depends on the masses of the new
particles, namely $m_{H^\pm}, m_{h^0}, m_{A^0}$, the leptonic
Yukawa couplings and total decay widths $\Gamma_{h^0},
\Gamma_{A^0}$. In our analysis, we observe large values, at the
order of magnitude of $10^{-4}\, GeV$, for the decay width of
the process, for outgoing $\tau$ and $\mu$ leptons. This is
informative in the determination of the upper limits of the Yukawa
couplings for LFV interactions and also in the prediction of the
new Higgs boson masses and the total decay widths of the new
neutral Higgs bosons.

We also analyze the LFC $H^+ \rightarrow W^+ l_i^- l_i^+$
$(l_i=\tau)$ decay in the model III. We observe that the decay
width of the process reaches to the values $10^{-3}\, GeV$,
depending on the appropriate choice of the free parameters. 
This analysis ensures a prediction for the leptonic constant, 
which is responsible for the $\tau-\tau$ transition.

The paper is organized as follows: In Section II, we present the
theoretical expression for the decay width of the LFV decay $H^+
\rightarrow W^+ l_i^- l_j^+$ and the LFC decay $H^+ \rightarrow
W^+ l_i^- l_i^+$, $l_i=\tau, l_j=\mu$, in the framework of the
model III. Section 3 is devoted to discussion and our conclusions.
%%%
%%%
\section{The charged Higgs $H^+ \rightarrow W^+ l_i^- l_j^+$  decay 
in the two Higgs doublet model}
In this section, we derive the expressions for the LFV $H^+
\rightarrow W^+ l_i^- l_j^+$ and LFC $H^+ \rightarrow W^+ l_i^-
l_j^+$ $(l_i=\tau, l_j=\mu)$ decays in the general 2HDM, the 
so-called model III. The leptonic part of the process can be 
regulated by the Yukawa interaction in the leptonic sector
\begin{eqnarray}
{\cal{L}}_{Y}= \eta^{E}_{ij} \bar{l}_{i L} \phi_{1} E_{j R}+
\xi^{E}_{ij} \bar{l}_{i L} \phi_{2} E_{j R} + h.c. \,\,\, ,
\label{lagrangian}
\end{eqnarray}
where $i,j$ are family indices of leptons, $L$ and $R$ denote
chiral projections $L(R)=1/2(1\mp \gamma_5)$, $\phi_{i}$ for
$i=1,2$, are the two scalar doublets, $l_{i L}$ and $E_{j R}$ are
lepton doublets and singlets respectively. On the other hand the
$H^+ \rightarrow W^+$ transition is possible with the help of the
scalar bosons,  the SM Higgs boson $H^0$ and CP even (odd) new
particle $h^0$ ($A^0$). The part of the lagrangian which is
responsible for these transitions is the so called kinetic term
\begin{eqnarray}
(D_{\mu} \phi_i)^+ D^{\mu} \phi_i = & & (\partial_{\mu} \phi_i^+ +
i\frac{g'}{2} B_{\mu} \phi^{+}_i+ i \frac{g}{2} \phi^+_{i}
\frac{\vec{\tau}}{2} \vec{W}_{\mu})
\nonumber \\
& & (\partial^{\mu} \phi_i - i\frac{g'}{2} B^{\mu} \phi_i- i
\frac{g}{2} \phi_{i} \frac{\vec{\tau}}{2} \vec{W}^{\mu}) .
\label{kinetic}
\end{eqnarray}
Here $\phi_{1}$ and $\phi_{2}$ are chosen as
\begin{eqnarray}
\phi_{1}=\frac{1}{\sqrt{2}}\left[\left(\begin{array}{c c}
0\\v+H^{0}\end{array}\right)\; + \left(\begin{array}{c c}
\sqrt{2} \chi^{+}\\ i \chi^{0}\end{array}\right) \right]\, ;
\phi_{2}=\frac{1}{\sqrt{2}}\left(\begin{array}{c c}
\sqrt{2} H^{+}\\ H_1+i H_2 \end{array}\right) \,\, ,
\label{choice}
\end{eqnarray}
where only $\phi_{1}$ has a vacuum expectation value;
\begin{eqnarray}
<\phi_{1}>=\frac{1}{\sqrt{2}}\left(\begin{array}{c c}
0\\v\end{array}\right) \,  \, ;
<\phi_{2}>=0 \,\, .
\label{choice2}
\end{eqnarray}
By considering the gauge and $CP$ invariant Higgs potential which
spontaneously breaks  $SU(2)\times U(1)$ down to $U(1)$  as:
\begin{eqnarray}
V(\phi_1, \phi_2 )&=&c_1 (\phi_1^+ \phi_1-v^2/2)^2+ c_2 (\phi_2^+
\phi_2)^2 \nonumber \\ &+&  c_3 [(\phi_1^+ \phi_1-v^2/2)+ \phi_2^+
\phi_2]^2 + c_4 [(\phi_1^+ \phi_1)
(\phi_2^+ \phi_2)-(\phi_1^+ \phi_2)(\phi_2^+ \phi_1)] \nonumber \\
&+& c_5 [Re(\phi_1^+ \phi_2)]^2 + c_{6} [Im(\phi_1^+ \phi_2)]^2
+c_{7} \, , \label{potential}
\end{eqnarray}
with constants $c_i, \, i=1,...,7$, $H_1$ and $H_2$ are obtained
as the mass eigenstates $h^0$ and $A^0$ respectively, since no
mixing occurs between two CP-even neutral bosons $H^0$ and $h^0$
in the tree level and the internal new scalars $h^0$ and $A^0$
play the main role for both $H^+ \rightarrow W^+ l_1^- l_2^+$ and
$H^+ \rightarrow W^+ l_1^- l_1^+$ decays (see Fig.1).

Now, we consider the lepton flavor changing process $H^+
\rightarrow W^+ l_i^- l_j^+$ where $l_i,\, l_j$ are different
leptons flavors, $e,\mu, \tau$. This process is drived by the
flavor changing (FC) interaction in the leptonic sector and the
strength of this interaction is carried by the Yukawa couplings
$\xi^{E}_{ij}$, which are the free parameters of the model III
version of 2HDM. They can have complex entries in general and
be restricted by using experimental measurements. Notice that, 
in the following, we replace $\xi^{E}$ with $\xi^{E}_{N}$ 
where "N" denotes the word "neutral".

The vertex function for $H^+ \rightarrow W^+$ is connected to the
$l_i^- l_j^+$ out going leptons by intermediate $h^0$ and $A^0$
bosons and the matrix element square of the process $H^+
\rightarrow W^+ l_i^- l_j^+$ is obtained as
\begin{eqnarray}
|M|^2&=& \frac{g^2}{2}\, h\, \{ \Bigg( \Big(
(m_{l_j}+m_{l_i})^2-k^2 \Big) |A^2|^2+\Big(
(m_{l_j}-m_{l_i})^2-k^2 \Big) |B^2|^2 \Bigg)|p_{h^0}|^2 \nonumber
\\  &+& \Bigg( \Big( (m_{l_j}+m_{l_i})^2-k^2 \Big) |A'^2|^2+\Big(
(m_{l_j}-m_{l_i})^2-k^2 \Big) |B'^2|^2 \Bigg)|p_{A^0}|^2 \nonumber \\
&-& 4 m_{l_j}\, m_{l_i}Im[(A\,A'^*-B\,B'^*)\,p_{h^0}\,p_{A^0}^*]
\nonumber\\ &-& 2 (m^2_{l_j}+m^2_{l_i}-k^2)\,
Im[(A\,A'^*+B\,B'^*)\,p_{h^0}\,p_{A^0}^*] \} \label{M2ij}
\end{eqnarray}
where
\begin{eqnarray}
h=\frac{k^2+(m^2_{H^\pm}-m^2_W)^2-2\, k^2
(m^2_{H^\pm}+m^2_W)^2}{m_W^2} \, , \label{hh}
\end{eqnarray}
and
\begin{eqnarray}
p_S=\frac{i}{k^2-m^2_S+i m_S\,\Gamma_S} \, . \label{pS}
\end{eqnarray}
with the transfer momentum square $k^2$. $\Gamma_S$ is the total
decay width of $S$ boson, for $S=h^0\,A^0$. In eq. (\ref{M2ij})
the factors $A, A', B, B'$ are the functions of the Yukawa
couplings;
\begin{eqnarray}
A&=&-\frac{i}{2\sqrt{2}}\, (\xi^E_{N,l_j l_i}+\xi^{* E}_{N,l_i
l_j})
\, , \nonumber  \\
A'&=&\frac{1}{2\sqrt{2}}\, (\xi^E_{N,l_j l_i}-\xi^{* E}_{N,l_i
l_j})\, , \nonumber  \\
B&=&-\frac{i}{2\sqrt{2}}\, (\xi^E_{N,l_j l_i}-\xi^{* E}_{N,l_i
l_j})
\, , \nonumber  \\
B'&=&\frac{1}{2\sqrt{2}}\, (\xi^E_{N,l_j l_i}+\xi^{* E}_{N,l_i
l_j}) \label{AB}
\end{eqnarray}
Finally, the decay width $\Gamma$ is obtained in the $H^\pm$ boson
rest frame using the well known expression
\begin{equation}
d\Gamma=\frac{(2\, \pi)^4}{m_{H^\pm}} \, |M|^2\,\delta^4
(p-\sum_{i=1}^3 p_i)\,\prod_{i=1}^3\,\frac{d^3p_i}{(2 \pi)^3 2
E_i} \,
 ,
\label{DecWidth}
\end{equation}
where $p$ ($p_i$, i=1,2,3) is four momentum vector of $H^+$ boson,
($W^+$ boson, incoming $l_j$, outgoing $l_i$ leptons).
%%%
%%%
\section{Discussion}
This section is devoted to the analysis of the charged Higgs
decays $H^+\rightarrow W^+\, (\tau^- \mu^+ + \tau^+ \mu^-)$ and
$H^+\rightarrow W^+\, \tau^- \tau^+$. The Yukawa couplings
$\xi^{E}_{N,\tau\mu}$ ($\xi^{E}_{N,\tau\tau}$) play the main role
in the leptonic part of the LFV $H^+\rightarrow W^+\, (\tau^-
\mu^+ + \tau^+ \mu^-)$ (LFC $H^+\rightarrow W^+\, (l_\tau^-
l_\tau^+)$) process. These couplings are free parameters of the
model used and they should be restricted by respecting the appropriate 
experimental measurements. The upper limit of the coupling
$\xi^{E}_{N,\tau\mu}$ has been predicted as $\sim 0.15$, by using
experimental result of anomalous magnetic moment of muon in
\cite{Erilano}.  However, the strength of the coupling
$\xi^{E}_{N,\tau \tau}$ is an open problem and waiting for new
experimental results in the leptonic sector. Furthermore, the
total decay widths of $h^0$ and $A^0$ are unknown parameters and
we expect that they are at the same order of magnitude of
$\Gamma_{H^0} \sim (0.1-1.0)\, GeV$, where $H^0$ is the SM Higgs
boson.

Notice that the couplings $\xi^{E}_{N,\tau \tau}$ and
$\xi^{E}_{N,\tau \mu}$, are complex in general and in the
following, we use the parametrization
\begin{equation}
\xi^{E}_{N,ij}= \sqrt{\frac{4 G_F}{\sqrt {2}}}
\bar{\xi}^{E}_{N,ij} \, , \label{ksipar}
\end{equation}
where $G_F=1.6637 \times 10^{-5} (GeV^{-2})$ is the fermi
constant. In our numerical calculations, we take $m_W= 80 \,GeV$.

At this stage, we would like to discuss the various charged Higgs
decays which are dominant and can be used in the
calculation of the BRs. The candidates for these decay
modes of the charged Higgs boson are $H^+\rightarrow W^+ h^0$,
$H^+\rightarrow \tau^+ \nu$ and $H^+\rightarrow
t\bar{b}$ \cite{Christove, Barger, Santos, Moretti}. The total 
decay width of the charged Higgs boson is approximated by
\begin{eqnarray}
\Gamma_{tot} (H^+)=\Gamma (H^+\rightarrow W^+ h^0)+\Gamma
(H^+\rightarrow  t\bar{b})+ \Gamma (H^+\rightarrow \tau^+ \nu)+
\Gamma (H^+\rightarrow c \bar{s}). \nonumber
\end{eqnarray}
Here we present the various BRs of the charged Higgs boson
decays:
\begin{eqnarray}
BR (H^+\rightarrow t\bar{b}) &<& 1 \nonumber \\
BR (H^+\rightarrow \tau^+ \nu)&<& 0.1 \nonumber \\
BR(H^+\rightarrow W^+ h^0) &<& 0.01 \nonumber \\
BR (H^+\rightarrow \mu^+ \nu)&<& 0.001 \nonumber \\
BR (H^+\rightarrow c\bar{s}) &<& 0.0001 \, , \label{NumBR1}
\end{eqnarray}
have been obtained,  for $tan\beta \sim 10$ and $m_{H^+}\sim 400
GeV$, in the  MSSM \cite{Moretti}. These results are strongly
sensitive to the choice of $tan\beta$, and increasing values of
$tan\beta$ make $H^+ \rightarrow \tau^+ \nu$ and $H^+ \rightarrow
\mu^+ \nu$ more dominant compared to the decay $H^+ \rightarrow
W^+ h^0$. In \cite{Diaz3}, $H^+ \rightarrow W^+ h^0$ has been
predicted at the order of $O(1)$ , in the context of the effective
lagrangian extension of the 2HDM.

Now we start to analyze the 3-body decay $H^+\rightarrow W^+\,
(\tau^- \mu^+ + \tau^+ \mu^-)$. In Fig.\ref{dWLFVmutau}, we present 
$\bar{\xi}^{E}_{N,\tau \mu}$ dependence of the decay width $\Gamma$ 
for the decay $H^+\rightarrow W^+\, (\tau^- \mu^+ + \tau^+ \mu^-)$, 
for the real coupling $\bar{\xi}^{E}_{N,\tau \mu}$,
$\Gamma_{A^0}=\Gamma_{h^0}=0.1 \,GeV$, $m_{h^0}=85\, GeV$ and
$m_{A^0}=90\, GeV$. Here solid (dashed, small dashed) line
represents the case for the Higgs mass $m_{H^\pm}=200 \, (300,
400) GeV$. The $\Gamma$ is strongly sensitive to the coupling
$\bar{\xi}^{E}_{N,\tau \mu}$, since it is proportional to square
of this coupling. Furthermore, this figure shows that the $\Gamma$
enhances with the increasing values of the charged Higgs mass, as
expected. The $\Gamma$ is at the order of magnitude of
$10^{-11}\, GeV$ for $m_{H^\pm}=200 \,GeV$ and it enhances to the
values $10^{-5}\, GeV$ for $m_{H^\pm}=400 \,GeV$, for even the
intermediate values of $\bar{\xi}^{E}_{N,\tau \mu}$. Fig.
\ref{dWLFVmHpl} represents the $m_{H^\pm}$ dependence of the $\Gamma$
for the fixed values of $\bar{\xi}^{E}_{N,\tau \mu}=1\, GeV$,
$\Gamma_{A^0}=\Gamma_{h^0}=0.1 \,GeV$, $m_{h^0}=85\, GeV$ and
$m_{A^0}=90\, GeV$. It is observed that the $\Gamma$ reaches large
values at the order of  magnitude of $10^{-5}$ even for the
small coupling $\bar{\xi}^{E}_{N,\tau \mu}=1\, GeV$. This is
interesting in the determination of the upper limit for the
charged Higgs mass $m_{H^\pm}$ and also the coupling
$\bar{\xi}^{E}_{N,\tau \mu}$.

In Fig. \ref{dWLFVGammah0} we present the total decay width
$\Gamma_{h^0}$ dependence of the decay width $\Gamma$ for
$\Gamma_{A^0}=\Gamma_{h^0}$, $\bar{\xi}^{E}_{N,\tau \mu}=1\, GeV$,
$m_{H^\pm}=400 \,GeV$, $m_{h^0}=85\, GeV$ and $m_{A^0}=90\, GeV$.
$\Gamma$ is sensitive to $\Gamma_{h^0}$ and decreases with its
increasing values.

Here, we will make the same analysis for the lepton conserving
process $H^+\rightarrow W^+\, \tau^- \tau^+$.
Fig. \ref{dWLFCtautau} denotes the $\bar{\xi}^{E}_{N,\tau \tau}$
dependence of the decay width $\Gamma$, for the real coupling,
$\Gamma_{A^0}=\Gamma_{h^0}=0.1 \,GeV$, $m_{h^0}=85\, GeV$ and
$m_{A^0}=90\, GeV$. Here solid (dashed, small dashed) line
represents the case for the mass value $m_{H^\pm}=200 \, (300,
400)\, GeV$. The $\Gamma$ is strongly sensitive to the coupling
$\bar{\xi}^{E}_{N,\tau \tau}$. It enhances with the increasing
values of the charged Higgs mass and it is placed in the interval
$10^{-9}-10^{-4}\, (GeV)$ for $200 (GeV)\leq
m_{H^\pm}\leq 400 (GeV)$, at the intermediate values of the
coupling $\bar{\xi}^{E}_{N,\tau \tau}$. In Fig. \ref{dWLFCmHpl},
we present the $m_{H^\pm}$ dependence of the $\Gamma$ for
$\bar{\xi}^{E}_{N,\tau \tau}=10\, GeV$,
$\Gamma_{A^0}=\Gamma_{h^0}=0.1 \,GeV$, $m_{h^0}=85\, GeV$ and
$m_{A^0}=90\, GeV$. From the figure it is seen that the $\Gamma$ 
reaches the large values at the order of magnitude of $10^{-4}$, 
even for the small coupling $\bar{\xi}^{E}_{N,\tau \tau}=10\, GeV$. 
The determination of the upper limit for the coupling
$\bar{\xi}^{E}_{N,\tau \tau}$ would be possible with the
measurement of the process under consideration.

Fig. \ref{dWLFCGammah0} represent $\Gamma_{h^0}$ dependence of the
decay width $\Gamma$ for $\Gamma_{A^0}=\Gamma_{h^0}$,
$\bar{\xi}^{E}_{N,\tau \tau}=10\, GeV$, $m_{H^\pm}=400 \,GeV$,
$m_{h^0}=85\, GeV$ and $m_{A^0}=90\, GeV$. $\Gamma$ is sensitive
to $\Gamma_{h^0}$ and decreases with its increasing values, similar 
to the LFV process $H^+\rightarrow W^+\,\tau^- \mu^+$.

Finally, we consider the coupling $\bar{\xi}^{E}_{N,l_i l_j}$
complex
\begin{equation}
\bar{\xi}^{E}_{N,l_i l_j}=|\bar{\xi}^{E}_{N,l_i l_j}|\, e^{i
\theta_{l_i l_j}} \, , \label{xicomplex}
\end{equation}
and study the $\sin\,{\theta_{l_i l_j}}$ dependence of the decay
width. We observe that the decay width is not sensitive to the
complexity of the coupling $\bar{\xi}^{E}_{N,l_i l_j}$.

At this stage we would like to summarize our results:

\begin{itemize}
\item We predict the decay width $\Gamma (H^+\rightarrow W^+\,
(\tau^- \mu^+ + \tau^+ \mu^-)$ ($\Gamma (H^+\rightarrow W^+\,
\tau^- \tau^+$) in the interval $(10^{-11}-10^{-5})\, GeV$
($(10^{-9}-10^{-4})\, GeV$), for $200 (GeV)\leq m_{H^\pm}\leq 400
(GeV)$, at the intermediate values of the coupling
$\bar{\xi}^{E}_{N,\tau \mu}\sim 5 \,GeV$ ($\bar{\xi}^{E}_{N,\tau
\tau}\sim \,30\, GeV$). With the possible experimental measurement
of the processes under consideration, strong clues would be
obtained in the prediction of the upper limit of the coupling
$\bar{\xi}^{E}_{N,\tau \mu}$ ($\bar{\xi}^{E}_{N,\tau \tau}$) .
This result is also informative in the determination of the
charged Higgs mass, $m_{H^\pm}$.

\item We observe that the decay width $\Gamma (H^+\rightarrow
W^+\, (\tau^- \mu^+ + \tau^+ \mu^-))$ ($\Gamma (H^+\rightarrow
W^+\, \tau^- \tau^+$) is strongly sensitive to the charged Higgs
mass, $m_{H^\pm}$.

\item We observe that the decay width $\Gamma (H^+\rightarrow
W^+\, (\tau^- \mu^+ + \tau^+ \mu^-))$ ($\Gamma (H^+\rightarrow
W^+\, (\tau^- \tau^+)$) is not sensitive to the possible
complexity of the Yukawa coupling.
\end{itemize}

Therefore, the experimental and theoretical analysis of these
decays of the charged Higgs boson would ensure strong hints in the
determination of the physics beyond the SM and the existing free
parameters.

\section{Acknowledgement}
This work has been supported by the Turkish Academy of Sciences in
the framework of the Young Scientist Award Program.
(EOI-TUBA-GEBIP/2001-1-8)

\newpage
\begin{figure}[htb]
\vskip 0.0truein \centering \epsfxsize=6.8in
\leavevmode\epsffile{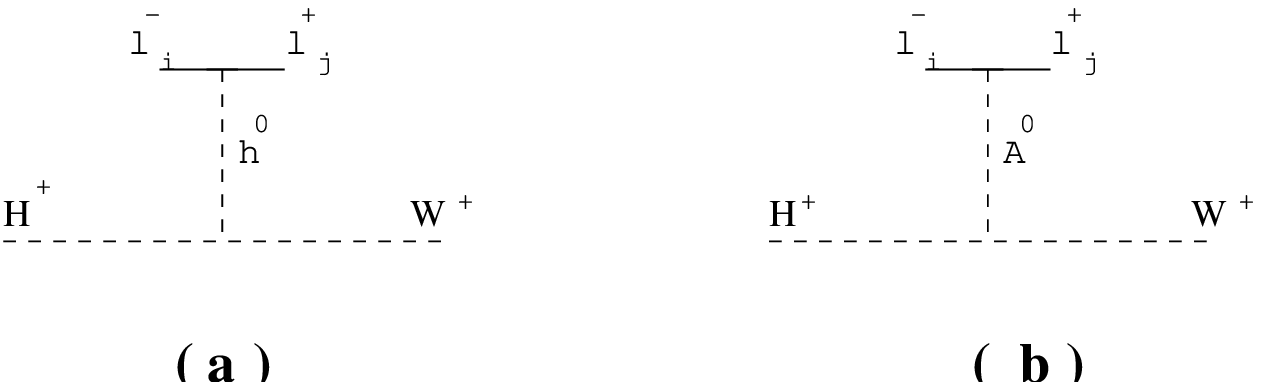} \vskip 1.0truein \caption[]{Tree
level diagrams contribute to $\Gamma (H^+\rightarrow W^+\, l_i^-
l_j^+)$, $i=e,\mu,\tau$ decay  in the model III version of 2HDM.
Solid lines represent leptons, dashed lines represent the $H^+,
W^+$, $h_0$ and $A_0$ fields).} \label{fig1}
\end{figure}
\newpage
\begin{figure}[htb]
\vskip -3.0truein \centering \epsfxsize=6.8in
\leavevmode\epsffile{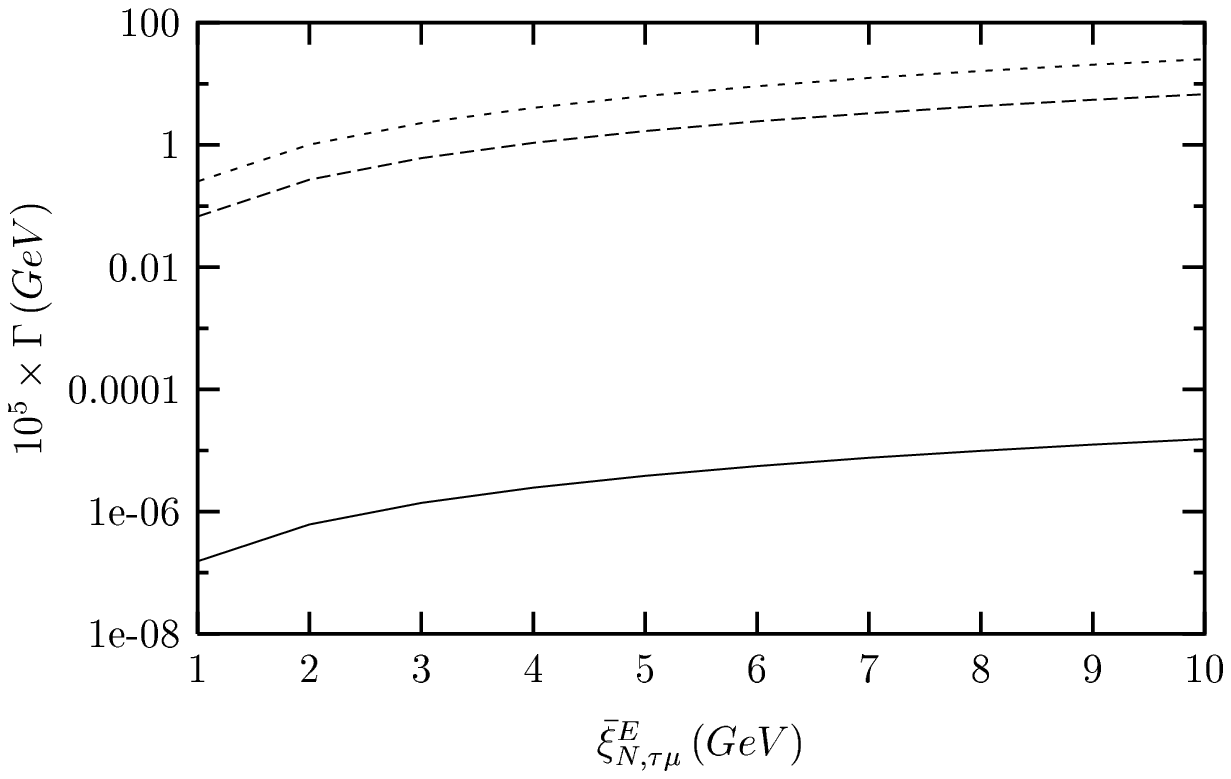} \vskip -3.0truein \caption[]{
$\bar{\xi}^{E}_{N,\tau \mu}$ dependence of the decay width $\Gamma
(H^+\rightarrow W^+\, (\tau^- \mu^+ + \tau^+ \mu^-))$, for the
real coupling $\bar{\xi}^{E}_{N,\tau \mu}$,
$\Gamma_{A^0}=\Gamma_{h^0}=0.1\, GeV$ $m_{h^0}=85\, GeV$ and
$m_{A^0}=90\, GeV$. Here solid (dashed, small dashed) line
represents the case for the mass value $m_{H^\pm}=200 \, (300,
400)\, GeV$.} \label{dWLFVmutau}
\end{figure}
\begin{figure}[htb]
\vskip -3.0truein \centering \epsfxsize=6.8in
\leavevmode\epsffile{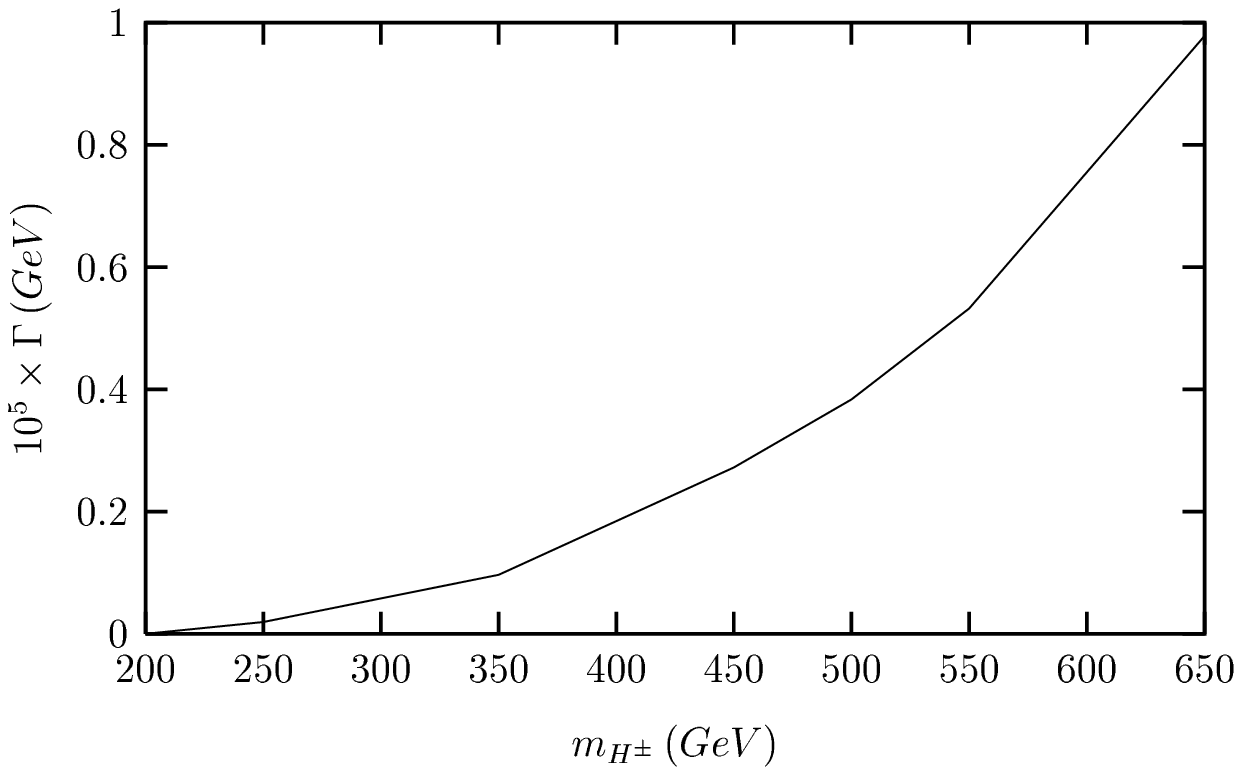} \vskip -3.0truein
\caption[]{The $m_{H^\pm}$ dependence of the decay width $\Gamma
(H^+\rightarrow W^+\, (\tau^- \mu^+ + \tau^+ \mu^-))$ for the
fixed values of $\bar{\xi}^{E}_{N,\tau \mu}=1\, GeV$,
$\Gamma_{A^0}=\Gamma_{h^0}=0.1\, GeV$, $m_{h^0}=85\, GeV$ and
$m_{A^0}=90\, GeV$.} \label{dWLFVmHpl}
\end{figure}
\begin{figure}[htb]
\vskip -3.0truein \centering \epsfxsize=6.8in
\leavevmode\epsffile{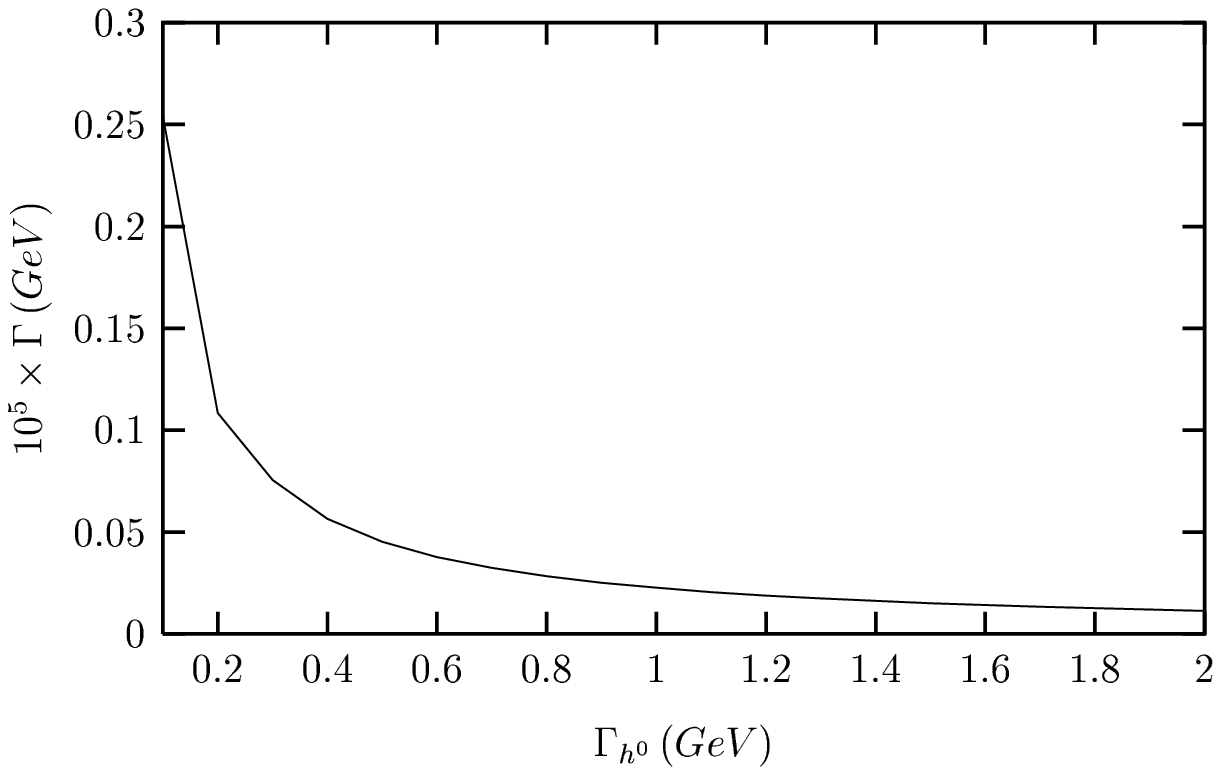} \vskip -3.0truein
\caption[]{ $\Gamma_{h^0}$ dependence of the decay width $\Gamma
(H^+\rightarrow W^+\, (\tau^- \mu^+ + \tau^+ \mu^-))$ for
$\Gamma_{A^0}=\Gamma_{h^0}$, $\bar{\xi}^{E}_{N,\tau \mu}=1\, GeV$,
$m_{H^\pm}=400 \,GeV$, $m_{h^0}=85\, GeV$ and $m_{A^0}=90\, GeV$.}
 \label{dWLFVGammah0}
\end{figure}
\begin{figure}[htb]
\vskip -3.0truein \centering \epsfxsize=6.8in
\leavevmode\epsffile{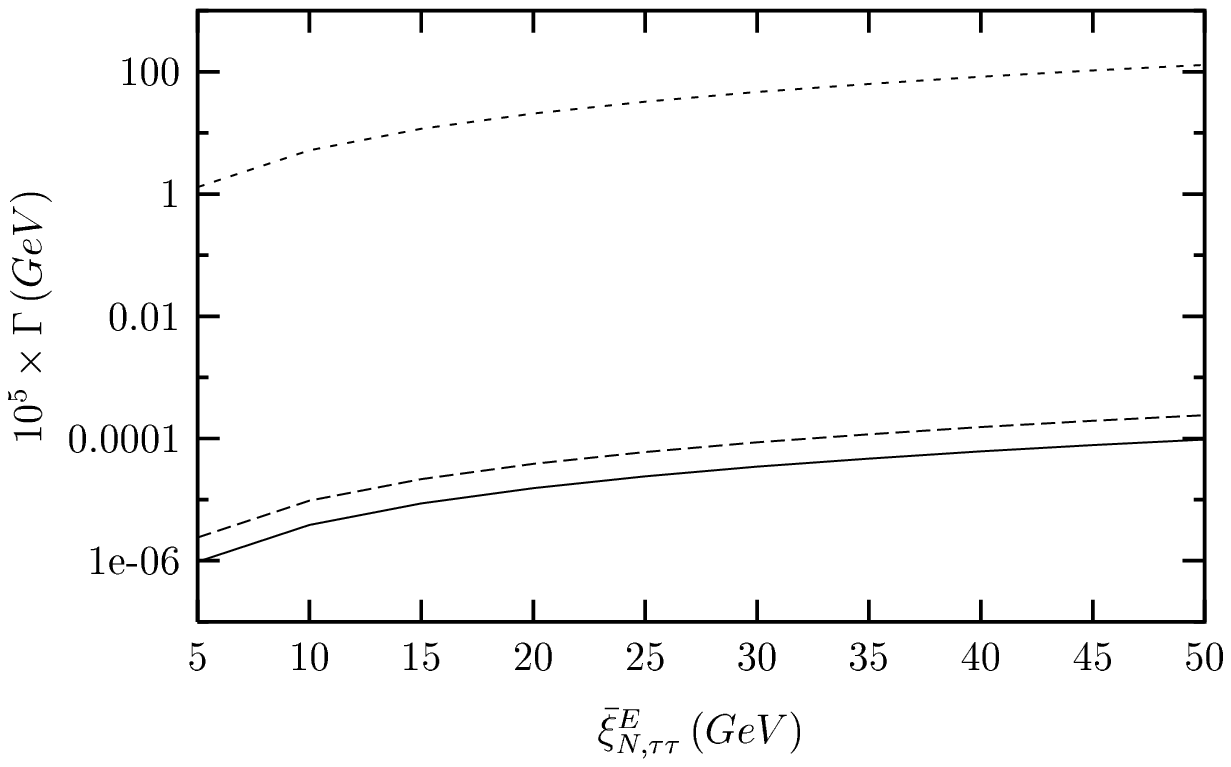} \vskip -3.0truein \caption[]{
$\bar{\xi}^{E}_{N,\tau \tau}$ dependence of the decay width
$\Gamma (H^+\rightarrow W^+\, \tau^- \tau^+)$, for the real
coupling $\bar{\xi}^{E}_{N,\tau \tau}$,
$\Gamma_{A^0}=\Gamma_{h^0}=0.1 \,GeV$ $m_{h^0}=85\, GeV$ and
$m_{A^0}=90\, GeV$. Here solid (dashed, small dashed) line
represents the case for the mass value $m_{H^\pm}=200 \, (300,
400)\, GeV$.} \label{dWLFCtautau}
\end{figure}
\begin{figure}[htb]
\vskip -3.0truein \centering \epsfxsize=6.8in
\leavevmode\epsffile{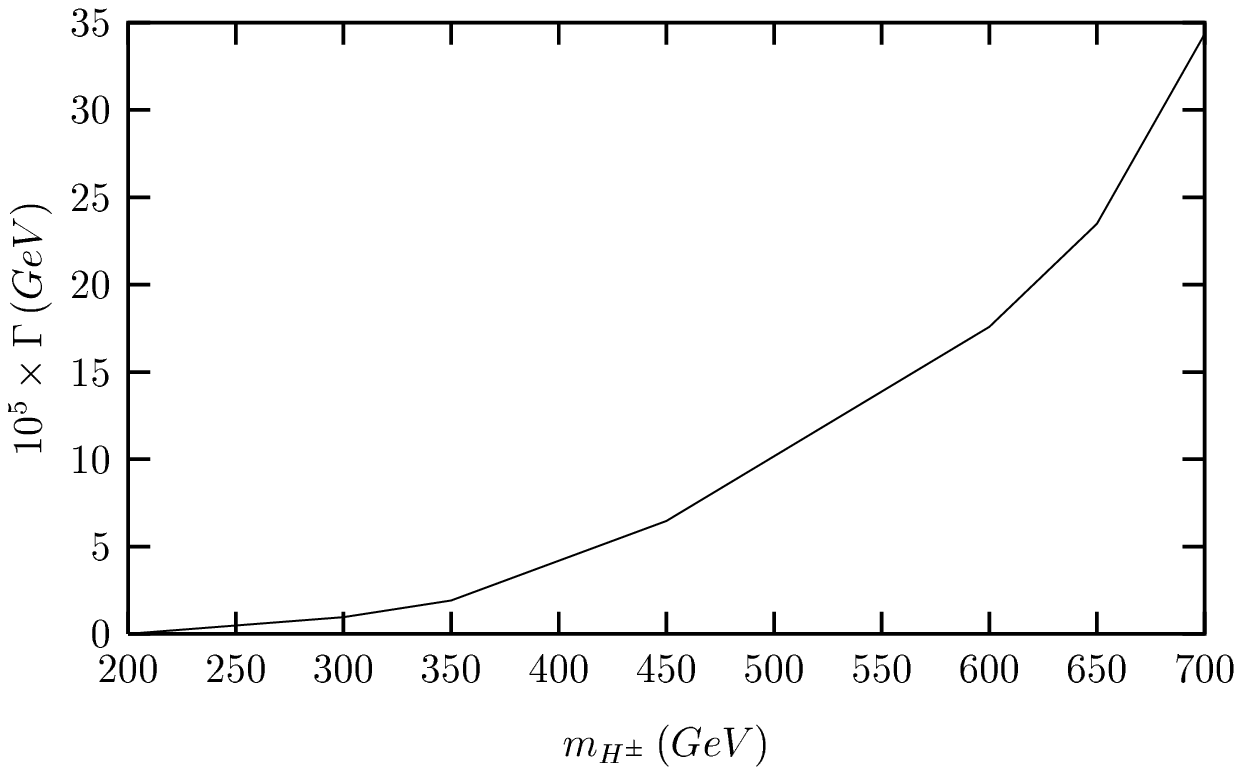} \vskip -3.0truein
\caption[]{The $m_{H^\pm}$ dependence of the decay width $\Gamma
(H^+\rightarrow W^+\, \tau^- \tau^+)$ for the fixed values of
$\bar{\xi}^{E}_{N,\tau \tau}=10\, GeV$,
$\Gamma_{A^0}=\Gamma_{h^0}=0.1\, GeV$, $m_{h^0}=85\, GeV$ and
$m_{A^0}=90\, GeV$.} \label{dWLFCmHpl}
\end{figure}
\begin{figure}[htb]
\vskip -3.0truein \centering \epsfxsize=6.8in
\leavevmode\epsffile{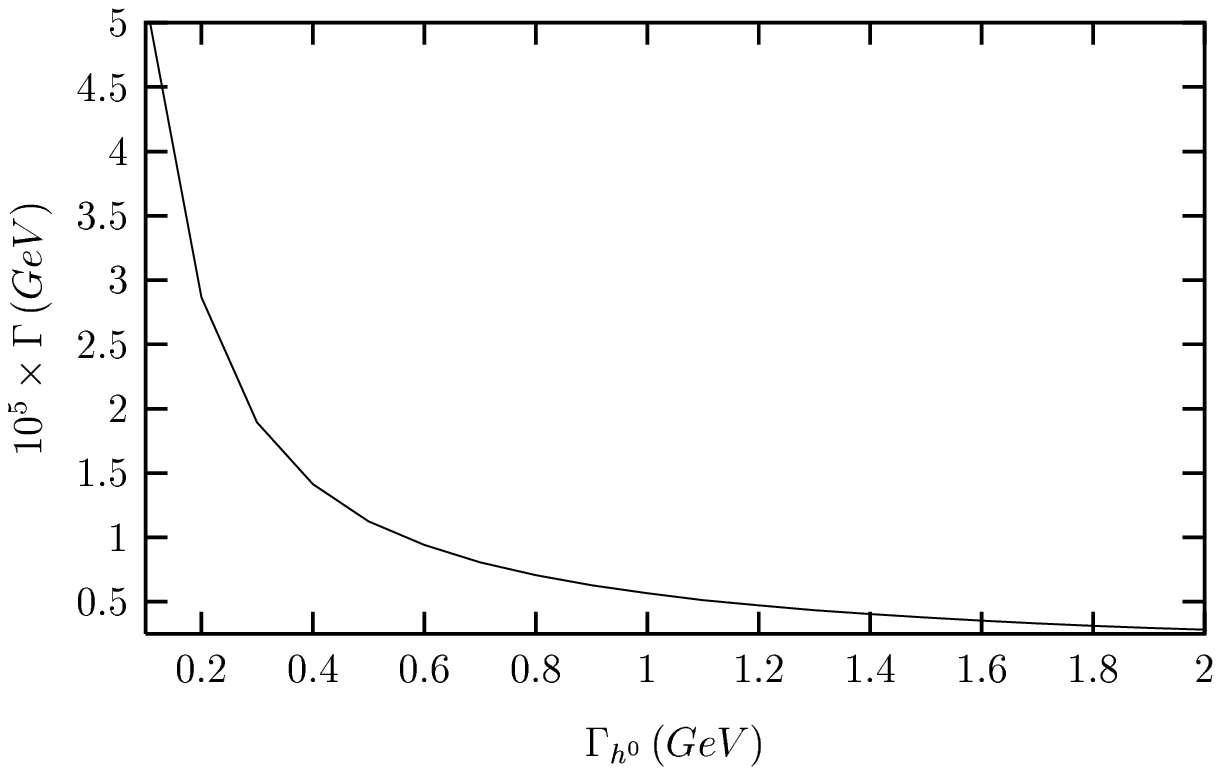} \vskip -3.0truein
\caption[]{ $\Gamma_{h^0}$ dependence of the decay width $\Gamma
(H^+\rightarrow W^+\, \tau^- \tau^+)$ for
$\Gamma_{A^0}=\Gamma_{h^0}$, $\bar{\xi}^{E}_{N,\tau \tau}=10\,
GeV$, $m_{H^\pm}=400 \,GeV$, $m_{h^0}=85\, GeV$ and $m_{A^0}=90\,
GeV$.}
 \label{dWLFCGammah0}
\end{figure}
\end{document}